\documentclass{llncs}
\usepackage{graphicx} 
\usepackage{booktabs} 
\usepackage{tabulary}
\usepackage{amsmath}
\usepackage{url}
\usepackage[shortlabels]{enumitem}
\usepackage{xcolor}
\usepackage{comment}

\newcommand{\ian}[1]{\textcolor{red}{ID: #1}}

\title{FASR: Automated Identification of Unsafe Control Actions in STPA}
\author{Ian Dardik\inst{1}\and
Yining She\inst{1}\orcidID{0000-0002-0071-501X} \and
Sam Procter\inst{2}\orcidID{0000-0003-4379-2362} \and
Keaton Hanna\inst{2}\orcidID{0000-0002-4434-0352} \and
Lutz Wrage\inst{2}\orcidID{0000-0003-4239-4768} \and
Eunsuk Kang\inst{1}\orcidID{0000-0001-7891-6885}}
\authorrunning{I. Dardik \etal}
\institute{School of Computer Science, Carnegie Mellon University, Pittsburgh PA 15213\\
\email{\{idardik,yiningsh,eunsukk\}@andrew.cmu.edu}\\
\and
Software Engineering Institute, Carnegie Mellon University, Pittsburgh PA 15213
\email{\{sprocter,kehanna,lwrage\}@sei.cmu.edu}}
\date{January 2026}

\begin{document}

\maketitle

\newcommand{\eg}{e.g.}
\newcommand{\ie}{i.e.}
\newcommand{\etc}{etc.}
\newcommand{\ibid}{ibid.}
\newcommand{\fig}{Figure\ }
\newcommand{\figs}{Figures\ }
\newcommand{\tbl}{Table\ }
\newcommand{\tbls}{Tables\ }
\newcommand{\sect}{Section\ }
\newcommand{\sects}{Sections\ }
\newcommand{\ftnt}{Footnote\ }
\newcommand{\defn}{Definition\ }
\newcommand{\lst}{Listing\ }
\newcommand{\alg}{Algorithm\ }
\newcommand{\pg}{page\ }
\newcommand{\etal}{et al.}
\newcommand{\itemify}[1]{\textbf{{#1}:}}

\begin{abstract} The \emph{System-Theoretic Process Analysis (STPA)} is a well-established hazard analysis technique that has been applied to a wide range of safety-critical systems. Despite its popularity, there is relatively little automation support for STPA, and most of its steps are carried out manually by a human analyst, which can be time consuming and error prone. This paper investigates the potential use of model-based engineering and formal methods to assist human analysts in efficiently and accurately carrying out STPA. The proposed tool, called \emph{FASR (Formalizing and Automating STPA with Robustness)}, enables automated, complete identification of \emph{unsafe control actions (UCAs)}, leveraging recent advances in \emph{robustness analysis} to identify UCAs as undesirable deviations in the controller's actions. The use of the tool is demonstrated on a case study involving a Braking System Control Unit (BSCU) in an avionics system. As a preliminary exploration of the potential benefits and limitations of the tool, the paper reports on a user study involving nine participants with varying backgrounds in STPA, model-based engineering, and formal methods; the study found that most participants considered the tool a useful aid in identifying UCAs, while suggesting improvements that would make a tool such as FASR usable and applicable to a wider range of systems and analysts. 
\end{abstract}

\section{Introduction}

The \emph{System-Theoretic Process Analysis (STPA)} is a well-established hazard analysis technique that has been applied to a wide range of safety-critical systems~\cite{leveson-thomas:BOOK18}. STPA, like other hazard analysis techniques, is a labor-intensive process: It requires a detailed understanding of the system structure and domain knowledge about potential hazards, and involves reasoning about possible interactions between different parts of the system, which can be a time-consuming and error-prone task. The amount of confidence that one has in the results of an STPA analysis (i.e., whether it has been carried out thoroughly and identified important hazards) is heavily dependent on the level of expertise and knowledge that the analyst possesses; thus, an analysis error or oversight might lead to an omission of a critical hazard scenario. While a tool cannot replace human knowledge or insight, one that supports the analyst in systematically enumerating hazards or checking for errors in their work may be highly beneficial and provide a higher level of confidence in the analysis output.



This paper investigates the use of model-based engineering and formal methods to provide tool support for STPA analysts. Our proposed tool, called \emph{FASR (Formalizing and Automating STPA with Robustness)}\footnote{\url{https://github.com/cmu-sei/fasr}}, specifically targets the automation of a crucial step in STPA that involves identification of \emph{unsafe control actions (UCAs)}; i.e., actions of a controller that result in its failure to enforce a desired safety constraint and possibly lead the system to a hazardous state. FASR accepts as input (1) a model of the system under analysis (including a controller and a set of processes that it controls), specified as state machines in SysML~\cite{omg-sysml:TECH24}, and (2) a safety invariant. As output, the tool automatically generates a list of UCAs that may cause the system to violate the invariant (i.e., transition into a hazardous state), to be examined by the human analyst for inclusion in the subsequent steps in STPA. Crucially, \textbf{FASR provides a formal guarantee that the list of UCAs is complete with respect to the details of the system as modeled}.

A key innovation underlying FASR is the adoption of recent advances in \emph{robustness analysis}~\cite{robustness-fse20,fortis-fmcad23}, a type of model-based, formal analysis used to analyze the ability of a system to remain safe under possible \emph{deviations} in the operating context. This analysis may be used to analyze, for example, whether or not an avionics braking system may enter a hazardous state when the human operator inadvertently deviates from their normative behavior, such as omitting an important action or performing actions out of order (i.e., a deviation). Later in the paper, we describe how the problem of identifying UCAs can be formulated as that of generating unsafe deviations and automated using an existing robustness analysis tool~\cite{fortis-fmcad23}.

To identify the potential benefits and limitations of FASR, we performed an exploratory user study involving nine participants with varying backgrounds in formal methods, model-based engineering, and STPA. In the study, participants were asked to use the tool to generate a list of UCAs for a hypothetical Braking System Control Unit (BSCU) in an avionics system; participants were then asked to manually evaluate the quality of the generated UCAs and determine whether they represented legitimate UCAs, sufficient to be included in a human-authored STPA report. Our findings indicate that most of the participants found the tool useful for identifying UCAs, especially those without prior experience with STPA, and increased their confidence in the final analysis output. The study also identified a number of challenges with the use of a model-based, automated tool like FASR, including (i) building a faithful and sufficiently detailed system model and (ii) interpreting the output of the formal analysis.



The contributions of the paper are as follows:
\begin{itemize}
\item An approach to automating the identification of STPA's UCAs through robustness analysis, 
\item A method for automatically classifying those UCAs according to ways that STPA specifies that control actions can be unsafe, \ie, its guidewords,
\item A prototype tool called FASR that implements this approach and classification, and, 
\item An exploratory user study involving the use of the FASR tool and a report on its findings, including perceived benefits and limitations of the FASR tool as well as future research directions.
\end{itemize}
\section{Background}
\subsection{STPA}
The \emph{System-Theoretic Process Analysis} is a well-established hazard analysis technique that views safety problems as resulting from inadequate control structures: it asks analysts to identify interactions in a critical system that could, if safety constraints are not enforced, lead to an accident or loss \cite{leveson-thomas:BOOK18}. Unlike more traditional, reliability-focused analyses (\eg, Failure Modes, Effects, and Criticality Analysis (FMECA) or Fault Tree Analysis (FTA) \cite{ericson:BOOK16}), its primary focus is not component failures. In addition to those, it also considers problems like operator errors, interface mismatches, as well as socio-technical concerns. The technique can be applied at different abstraction levels, ranging from organizational to human-computer interaction to low-level automation \cite{leveson-thomas:BOOK18}. 

STPA has four steps: in the first step, the analyst documents the system, its environment, and the boundary between them. The analyst also describes the losses that can occur in the environment due to the system, as well as the hazards which would cause those losses and safety constraints which would prevent them. In the second step, the analyst models the control structure, including the control actions that controller components send to controlled processes. In the third step, the analyst considers how those control actions could be unsafe according to several \emph{guidewords}, such as \emph{providing} (\ie, if the control action is provided when it should not be) or \emph{not providing} (\ie, if the control action is not provided when it should be). The fourth step requires the analyst to consider if and how the UCAs identified in the third step could occur given the system's implementation.

\subsection{Model-Based Engineering}
In this paper, we present our approach on a \emph{model} of a system and its components.
Analyzing a system model---rather than its implementation---is advantageous for reducing the complexity of the analysis as well as ensuring that the system design is safe before implementation.
More generally, the practice of \emph{Model-Based Engineering} (MBE) is well-established in the development of critical systems, where the ``model'' may range from an informal sketch on a whiteboard to a detailed description written in a domain-specific language with precise semantics.
A wide range of MBE languages and tools exist for a similarly wide range of applications \cite{de_saqui-sannes_taxonomy_2022}, including the use of those models for safety analysis.
One of the most popular languages for MBE is SysML \cite{de_saqui-sannes_taxonomy_2022}, which contains a variety of diagram types for modeling different aspects of a system's architecture, usage, and functionality \cite{omg-sysml:TECH24}.

\textbf{Model-Based Safety Assessment (MBSA).}
MBE has been incorporated into safety assessments in various ways, including for manual hazard analyses (a human analyst references a system model) and tool-assisted hazard analyses (e.g.,  report generation \cite{Procter2014}, change impact and traceability support \cite{ahlbrecht_model-based_2022}, and automated analysis of critical subsystems as described in this work).


\textbf{RAAML.}
The \emph{Risk Analysis and Assessment Modeling Language} (RAAML) is an extension of SysML for documenting system hazards and other safety information \cite{omg-raaml:TECH24}. It provides profiles and libraries to support risk / safety analysis. These include a library of core elements and method-specific extensions for different hazard analysis techniques including STPA. RAAML support---which includes editing and rendering support, but no automated analysis---is installable as a plugin in some modeling tools, such as Dassault Syst\`{e}mes' Cameo Enterprise Architecture (CEA) \cite{cameo2024x}. We implemented our approach using CEA and RAAML in order to more deeply integrate it into existing model-based tooling that practitioners are already using.

\vspace{-.75em}
\subsection{Robustness Analysis}
\label{sec:robustness-background}
\emph{Robustness} is a quality attribute that describes the ability of a system to handle external disturbances or deviations in the environment. In a recent line of work, Zhang et al.~\cite{robustness-fse20} have proposed a formal definition of robustness for computer systems and a model-based analysis tool, called \emph{Fortis}~\cite{fortis-fmcad23}, that can be used to analyze the robustness of a system. In particular, Fortis takes as input: (1) a model of a computer system ($M$), specified as a transition system, (2) a model of the normative environment ($E$), also specified as a transition system, and (3) a desired safety property ($P$); the tool then computes \emph{unsafe environmental deviations} that can lead to a violation of the safety property. In this setting, a \emph{deviation} is a \emph{trace} (i.e., a sequence of events in  $E$) that shows how the environment may behave differently from its normative behavior. For example, $M$ may describe the design of an avionics braking system; $E$ the expected behavior of a human operator; and $P$ a safety property stating that the brake is activated as the plane lands. In this example, an unsafe deviation would represent an erroneous sequence of actions by the operator (e.g., skipping or performing critical actions out-of-order) that could cause the plane to land without proper braking. 

Later in Section~\ref{sec:robustness-analysis}, we describe how the problem of finding UCAs can be formulated as generating unsafe deviations and given automated analysis support using a tool like Fortis.
\section{Case Study: BSCU Analysis with FASR}
\label{sec:bscu}

We selected a simplified avionics Braking System Control Unit (BSCU) system to help illustrate and evaluate our work.
The system is derived from a well-studied example system described in, among other publications, the STPA Handbook \cite{leveson-thomas:BOOK18}.
This case study captures a scenario in which a flight crew operates the BSCU of an airplane during a landing.
The flight crew is expected to follow the operating instructions that appear in the manual for the BSCU.
Correctly operating the BSCU is crucial for avoiding the key hazard in this case study: the airplane overshooting the runway during the landing due to improper braking.

\begin{figure}[!t]
\includegraphics[width=\textwidth]{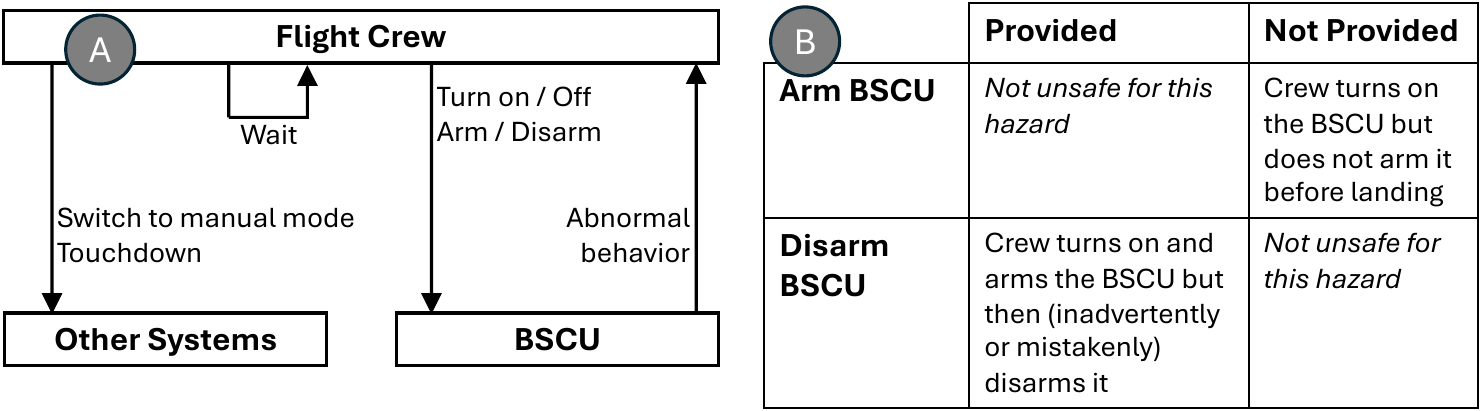}
\caption{(A) Control structure and sample (B) unsafe control actions for the Braking System Control Unit (BSCU) case study}
\label{fig:bscu-stpa-sample}
\end{figure}

\begin{figure}[!t]
\includegraphics[width=\textwidth]{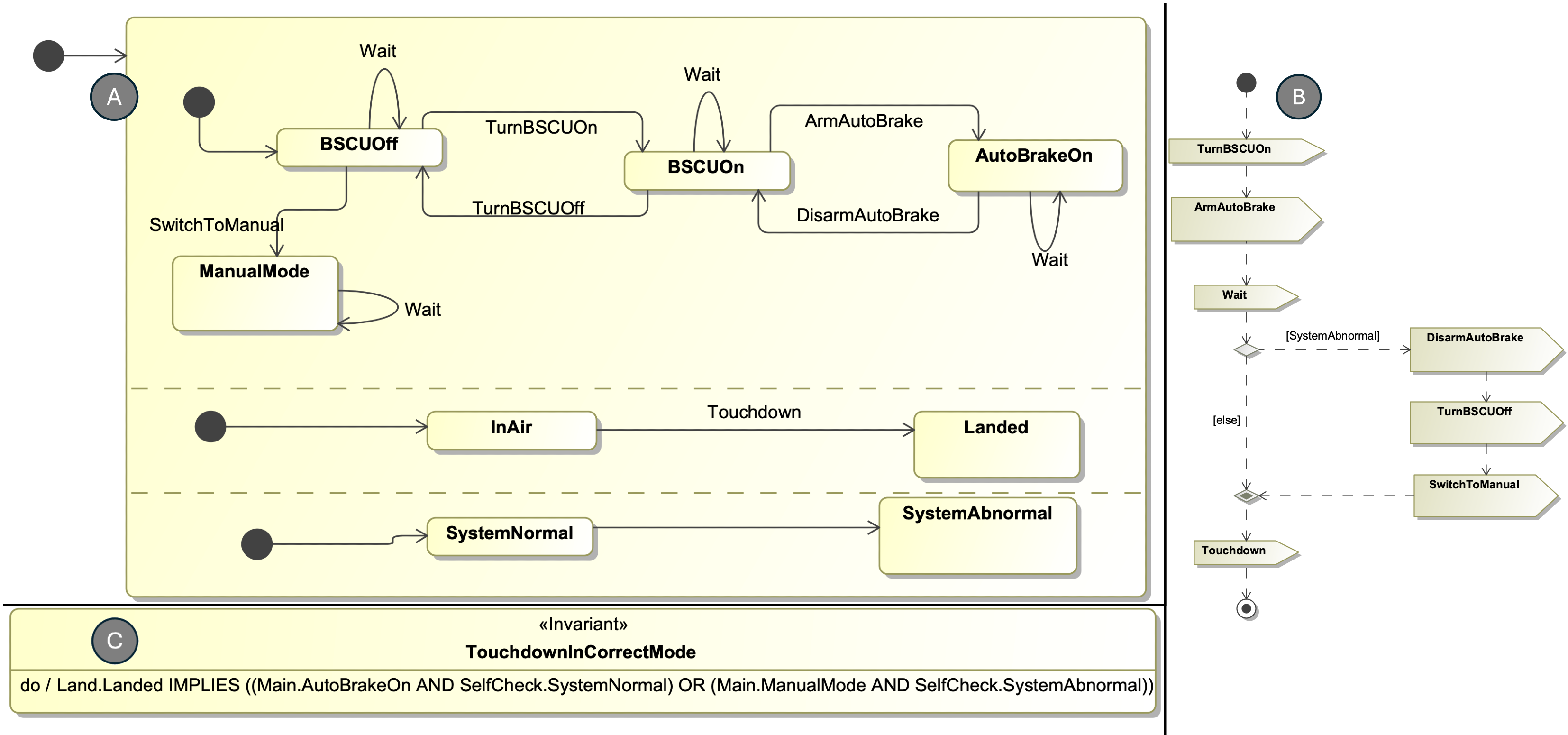}
\caption{A screenshot of Cameo Enterprise Architecture (CEA) showing the BSCU case study modeled using the SysML profile created for this effort. (A) is a state machine diagram of the BSCU, (B) is an activity diagram of the crew instructions, (C) is the hazard we want to avoid, expressed as an invariant.}
\label{fig:bscu-sysml}
\end{figure}

\textbf{STPA.}
In this case study, the flight crew represents the controller for the BSCU.
We show a diagram of the control structure in \fig\ref{fig:bscu-stpa-sample}(A), including actions that allow the flight crew to turn the BSCU on or off, arm or disarm the BSCU, and detect abnormalities in the baking unit.
In this case study, we will primarily focus on UCA identification, i.e. identifying control actions that lead to the airplane overshooting the runway.
We show a sample of the UCAs for this case study in \fig\ref{fig:bscu-stpa-sample}(B).
For example, the upper right-hand box in \fig\ref{fig:bscu-stpa-sample}(B) identifies that a hazard may occur if the flight crew fails to arm the BSCU before landing, which is required by the BSCU instruction manual.

\textbf{UCA Identification with FASR.}
Ordinarily, an STPA analyst must manually identify the UCAs in the third step.
In this paper, we advocate assisting UCA identification with an automated tool such as FASR.
A key advantage to this approach is that FASR is \emph{guaranteed to identify all UCAs} with respect to user-provided models.
This completeness guarantee is important because the control structure may be complex, difficult to reason about, and possibly tough for a human to identify all UCAs.

As input, FASR requires a model of the control structure in SysML.
We show the model for the BSCU in \fig\ref{fig:bscu-sysml}(A), the correct operating behavior of the flight crew (as specified in the BSCU manual) in \fig\ref{fig:bscu-sysml}(B), and the key safety requirement---an invariant that implies the airplane has not overshot the runway---in \fig\ref{fig:bscu-sysml}(C).
FASR then \emph{automatically} identifies all UCAs for the given model, which a human can then use to create a table similar to the one shown in \fig\ref{fig:bscu-stpa-sample}(B).

We emphasize that FASR's completeness guarantee is contingent on the correctness of the models of the control structure that are given as input.
In essence, our approach shifts an engineer's concern from completeness of UCAs to properly constructing a control structure model; this type of shift is common in MBE.

\section{FASR Tool}
\label{sec:tool}

\begin{figure}[!t]
\includegraphics[width=\textwidth]{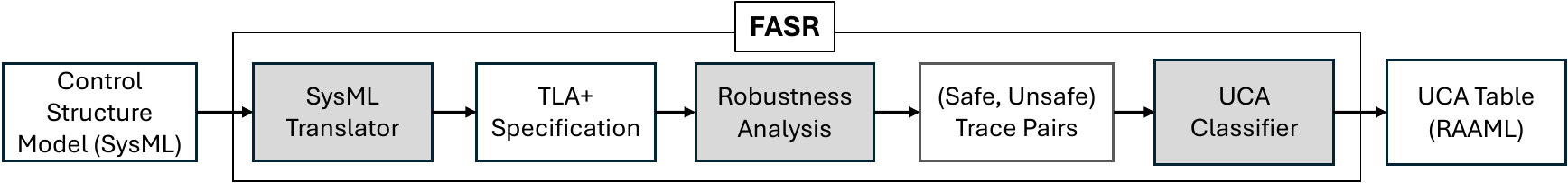}
\caption{An overview of the FASR tool.}
\label{fig:fasr}
\end{figure}

The high-level overview of the FASR tool is illustrated in Figure~\ref{fig:fasr}. The tool takes as input a model of the control structure written in SysML and outputs a list of UCAs that may cause the system to enter a hazardous state in RAAML. The rest of this section describes the main components of the FASR tool (highlighted in gray in Figure~\ref{fig:fasr}).

\subsection{SysML to TLA+ Translation}
\label{sec:translation}

The input SysML model consists of three different parts: 
\begin{itemize}
\item \emph{Controlled process}, specified as a state machine. The states of the machine are represented using a set of state variables, and each transition in the machine corresponds to an event that modifies one or more of these variables.  
\item \emph{Controller}, specified as an activity diagram that describes normative sequences of control actions. 
\item \emph{Invariant}, specified as a Boolean condition over  state variables, indicating the absence of a hazard to be avoided. The invariant is assumed to hold under the normative control sequences as specified in the controller model. 
\end{itemize}
The SysML model is then translated into a specification in the TLA+ language~\cite{lamport2002specifying}. TLA+ is a formal specification language that has been widely used for modeling and verifying concurrent and distributed computer systems. We omit the detailed translation scheme from SysML to TLA+ here but briefly, both the controller and the controlled process in SysML are translated into transition systems in TLA+, and invariants are translated into safety properties (based on temporal logic~\cite{ltl}) in TLA+.

\subsection{Unsafe Trace Generation through Robustness Analysis}
\label{sec:robustness-analysis}

In the next step, FASR leverages robustness analysis (implemented in the Fortis tool~\cite{fortis-fmcad23}) to generate traces that show how UCAs in the controller can lead the controlled process to a hazardous state. The key idea in this step is to treat the controller as an entity that may sometimes deviate from its normative behavior, and UCAs as those deviations that lead to a violation of the safety invariant.

Recall from Section~\ref{sec:robustness-background} that Fortis accepts models of the system ($M$) and the environment ($E$) and a safety property ($P$) as input, and produces a set of unsafe environmental deviations. In the context of FASR, the controlled process is treated as the system ($M$), the controller as the environment ($E$) and the invariant as the safety property ($P$). 
In particular, the transition system for $E$ in TLA+ encodes traces that correspond to normative control sequences; then, an unsafe deviation, generated by Fortis, is a trace that results in a violation of the safety invariant. 

Some readers might find it counter-intuitive at first that the controller, not the controlled process, is treated as the environment. Conceptually, in this setting, robustness analysis is being used to analyze how robust or brittle the controlled process (e.g., the BSCU) is against possible errors or mishaps in the controller (e.g., the flight crew deviating from ideal instructions).

In its output, Fortis accompanies each unsafe deviation with a safe trace that avoids the deviation.
Therefore, Fortis outputs trace \textit{pairs} of the form (\texttt{safe}, \texttt{unsafe}).
For example, consider the trace: $\langle \texttt{TurnBSCUOn}, \texttt{Wait}, \texttt{Touchdown}\rangle$.
This trace is unsafe because the flight crew fails to arm the BSCU before the airplane touches down, which may result in a hazard (the airplane overshooting the runway).
In the robustness setting, the trace is viewed as a deviation from the normative environment, which expects the flight crew to correctly arm the BSCU before touchdown as described in the instruction manual.
In contrast, the following trace is safe and expected from the normative environment:
$\langle \texttt{TurnBSCUOn}, \texttt{ArmAutoBrake}, \texttt{Wait}, \texttt{Touchdown}\rangle$.
Fortis would accompany this safe trace with the prior deviated trace in its output.
In the following section, we will describe how these pairs can be used to classify each UCA by an STPA guideword.

\subsection{UCA Classification}
\label{sec:uca-classification}

\begin{table*}[t]
    \begin{center}
        \begin{tabulary}{\columnwidth}{Llclclllr}
            \toprule
            Edit & \phantom{m} & \multicolumn{3}{c}{Removed} & \phantom{m} & Guideword & \phantom{m} & Control Action \\
            \cmidrule(){1-1} \cmidrule(){3-5} \cmidrule(){7-7} \cmidrule(){9-9}
            Deletion & \phantom{i} & \multicolumn{3}{c}{\texttt{Wait}} & \phantom{i} & Too Early & \phantom{i} & Early Action\\
            Deletion & \phantom{i} & \multicolumn{3}{c}{\texttt{Any}} & \phantom{i} & Not Providing & \phantom{i} & Removed Action\\
            \\
            Edit & \phantom{i} & \multicolumn{3}{c}{Added} & \phantom{i} & Guideword & \phantom{i} & Control Action \\
            \cmidrule(){1-1} \cmidrule(){3-5} \cmidrule(){7-7} \cmidrule(){9-9}
            Addition & \phantom{i} & \multicolumn{3}{c}{\texttt{Wait}} & \phantom{i} & Too Late & \phantom{i} & Late Action\\
            Addition & \phantom{i} & \multicolumn{3}{c}{\texttt{Any}} & \phantom{i} & Providing & \phantom{i} & Additional Action\\
            \\
            Edit & \phantom{i} & Correct & \phantom{m} & Incorrect & \phantom{i} & Guideword & \phantom{i} & Control Action \\
            \cmidrule(){1-1} \cmidrule(){3-3} \cmidrule(){5-5} \cmidrule(){7-7} \cmidrule(){9-9}
            Substitution & \phantom{i} & \texttt{Any} & \phantom{i} & \texttt{Any} & \phantom{i} & Providing & \phantom{i} & Provided Action \\
            Substitution & \phantom{i} & \texttt{Any} & \phantom{i} & \texttt{Wait} & \phantom{i} & Not Providing & \phantom{i} & Removed Action \\
            Substitution & \phantom{i} & \texttt{Wait} & \phantom{i} & \texttt{Any} & \phantom{i} & Providing & \phantom{i} & Additional Action \\
            Transposition & \phantom{i} & \texttt{Any} & \phantom{i} & \texttt{Any} & \phantom{i} & Out of Order & \phantom{i} & Incorrect Action \\
            Substitution & \phantom{i} & \texttt{Any} & \phantom{i} & \texttt{Wait} & \phantom{i} & Too Late & \phantom{i} & Late Action \\
            Substitution & \phantom{i} & \texttt{Wait} & \phantom{i} & \texttt{Any} & \phantom{i} & Too Early & \phantom{i} & Early Action \\
            \bottomrule
        \end{tabulary}
    \end{center}
    \caption{Mapping from atomic string edits in the Damerau-Levenshtein's edit distance algorithm \cite{damerau-distance} to STPA's Step 3 Guidewords. The \texttt{Any} action signifies any action other than \texttt{Wait} or the expected action from the safe trace (in the Removed, Added, or Correct columns) or unsafe trace (in the Incorrect column).}
    \label{tbl:dl-uca-mapping}
\end{table*}

The next and final step in FASR is to classify unsafe control sequences generated from the preceding robustness analysis into categories of UCAs that correspond to guidewords in STPA's third step. For each pair of safe and unsafe traces generated by Fortis, the UCA classifier selects a guideword that best explains the difference between the two traces. To do so, the classifier uses a modified version of the \emph{Damerau-Levenshtein} \cite{damerau-distance} algorithm. The unmodified algorithm computes the edit distance between two strings, \ie, the minimum number of atomic operations (\ie, addition, deletion, substitution of a character or transposition of two adjacent characters) required to transform one string into the other; our modifications were to make the algorithm operate on traces (rather than strings) and to map the atomic string edits to STPA's guidewords using the mapping in \tbl\ref{tbl:dl-uca-mapping}. 

For example, consider the following (\texttt{safe}, \texttt{unsafe}) trace pair:
\begin{align*}
(\langle \texttt{TurnBSCUOn}, \texttt{ArmAutoBrake}, \texttt{Wait}, \texttt{Touchdown}\rangle,
\langle \texttt{TurnBSCUOn}, \texttt{Wait}, \texttt{Touchdown}\rangle)
\end{align*}
The operation needed to transform the safe trace into its unsafe counterpart is the deletion of $\texttt{ArmAutoBrake}$ action; the classifier then selects the STPA guideword \emph{not providing} for this UCA. In other words, this UCA describes a scenario where the flight crew fails to arm the BCSU before landing. 

The guidewords \emph{provided too early} and \emph{too late} describe an incorrect timing of a control action being performed. Since SysML models or TLA+ specifications do not have a built-in notion of time, FASR designates a special action called \texttt{Wait} to denote passing of a unit time. The number of \texttt{Wait} actions in a trace is then used to determine whether a certain UCA is performed too early or too late. As another example, consider the following (\texttt{safe}, \texttt{unsafe}) trace pair: 
\begin{align*}
(\langle \texttt{ArmAutoBrake}, \texttt{Wait}, \texttt{Touchdown}\rangle,
\langle  \texttt{ArmAutoBrake},\texttt{Wait}, \texttt{Wait}, \texttt{Touchdown}\rangle)
\end{align*}
Since the unsafe trace contains additional \texttt{Wait}, signifying that the flight crew waits longer than expected before performing touchdown, this UCA is classified under the guideword \emph{too late} (for the \texttt{Touchdown} action). 

Note that the guidewords \emph{Applied too Long} and \emph{Stopped too Soon} are calculated using a separate mechanism: rather than using the modified Damerau-Levenshtein algorithm, users can specify actions in SysML which signify the start or end of ongoing processes. If the unsafe trace has more or fewer \texttt{Wait} actions between these start and stop actions than the safe trace, we classify the UCA as \emph{Applied too Long} and \emph{Stopped too Soon}, respectively.

While other automated UCA identification approaches exist (see, \eg, \cite{thomas2013formalModelBasedSafetyReqs,petzold_hot_2026} and others in \sect\ref{sec:related}), to our knowledge, the mapping of the problem to that of computing string edits and the use of a modified edit-distance calculation algorithm is novel.

\subsection{FASR Analysis of BSCU}

\begin{figure}[t]
\includegraphics[width=\textwidth]{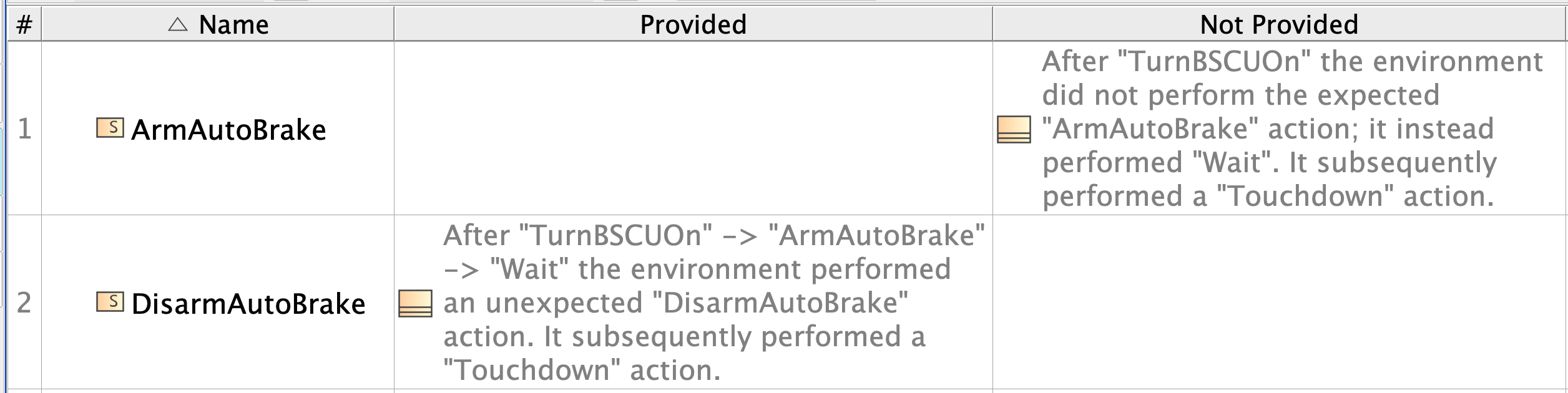}
\caption{An excerpt of the UCA table generated by FASR for the BSCU example.}
\label{fig:bscu-fasr-output}
\end{figure}

We ran the FASR toolchain on the BSCU example using an Apple M3 Max running macOS 26.3.1; some performance statistics are shown in \tbl\ref{tbl:perf}. Though performance wasn't a primary goal of this effort, the runtimes are likely to be acceptable: roughly 4-5 seconds when abnormal BSCU behavior was not considered and 6-8 seconds when it was. The state space explored by Fortis was significant: thousands of states were found when abnormal behavior was disregarded, and tens of thousands were found when it was included. \tbl\ref{tbl:perf} also shows the significant number of near-duplicate UCAs being discovered: more than half of Fortis-generated UCAs were filtered out by the FASR front-end, which performs only rudimentary de-duplication by, \eg, removing UCAs that are identical except in the number of consecutive wait actions. A screenshot of the generated UCA table (formatted using RAAML and displayed in Cameo Enterprise Architecture) is shown in \fig\ref{fig:bscu-fasr-output}; these UCAs correspond to those shown in \fig\ref{fig:bscu-stpa-sample}.

\begin{table*}[t]
    \begin{center}
        \begin{tabulary}{\columnwidth}{Lccccccccc}
            \toprule
            \phantom{i} & \phantom{i} & \multicolumn{2}{c}{Time (sec)} & \phantom{i} & \multicolumn{2}{c}{Generated UCAs} & \phantom{i} & \multicolumn{2}{c}{Fortis} \\
            \phantom{i} & \phantom{i} & Cold & Warm & \phantom{i} & Fortis & Filtered & \phantom{i} & States & Trans.\\
            \cmidrule(){3-4} \cmidrule(){6-7} \cmidrule(){9-10}
            BSCU, No Abnormal Behavior & \phantom{i} & 4.7 & 4.1 & \phantom{i} & 31 & 14 & \phantom{i} & 241 & 3100\\
            BSCU, With Abnormal Behavior & \phantom{i} & 7.5 & 5.8 & \phantom{i} & 55 & 23 & \phantom{i} & 1249 & 50024\\
            \bottomrule
        \end{tabulary}
    \end{center}
    \caption{Runtime (with cold and warm starts), no.\ of UCAs (generated by Fortis and shown to users), and no.\ of Fortis-explored states and transitions for  BSCU.}
    \label{tbl:perf}
\end{table*}

\section{Exploratory Study}
In this section, we present a user study in which participants analyzed the BSCU example using STPA with the aid of FASR. The study was exploratory in nature; our goal was to gather qualitative feedback to understand whether our approach provides benefits as well as how the tooling might be improved in future iterations. In particular, we study the following two research questions: 
\begin{enumerate}
    \item[] RQ1: Do study participants perceive any benefits from using FASR to identify UCAs?
    \item[] RQ2: What downsides and improvements to the tool do study participants identify?
\end{enumerate}

\textbf{Cohort Demographics} We invited ten participants from Carnegie Mellon University's Software Engineering Institute to participate in the study. Nine participants completed the study, of which six were given access to the tooling and three asked to perform STPA manually. Participants were asked to rate their familiarity with STPA, MBE, and formal methods; while most participants were familiar with MBE, there was a wider range of knowledge in STPA and formal methods.

\subsection{User Study Description}
In order to provide a baseline level of knowledge in the relevant techniques, we developed a one-hour training course which introduced STPA, the BSCU system as described in \sect\ref{sec:bscu}, and the tool itself (including the modeling environment it is a plugin for, Cameo Enterprise Architecture). The slides and a recording of the training were provided to participants afterwards for review.

We then met individually with each participant for an hour and had them examine the system models and perform the third step of STPA on the BSCU, \ie, identify which control actions in the scenario could be unsafe. The manual cohort filled out an UCA table directly. The tool cohort was allowed to use the FASR software to generate candidate UCAs, but was asked to evaluate their quality (to determine if they were legitimate or spurious) and rephrase or rewrite them as much as necessary so they would be clear to a third party reviewer. The tool cohort was also encouraged to record additional UCAs they thought of that the tool did not generate. 

Third, we interviewed each participant individually to collect qualitative data on the FASR tool. We asked participants a series of demographic and open-ended questions about their overall experience using it to evaluate the BSCU, the tool's strengths and weaknesses, as well as any potential improvements.

\subsection{Results and Discussion}
\emph{\textbf{RQ1:}}
In response to the question ``Would you use this tool for identifying hazards in safety-critical systems?'' every participant answered in the affirmative, though some more strongly than others. One participant, who had previously worked with different tool support for STPA, was particularly enthusiastic, responding: ``This is absolutely the sort of tool I would use [if I were still working in this area]. I really like what I'm seeing here, [the tool] generated more information than the tools we used to have that were less connected to formalisms.'' Another participant, however, responded to the same question with ``Yeah, [although] this iteration, maybe not yet\ldots\ This would be super helpful if you could clean it up a bit.''

By giving them a starting point to work from, rather than a blank page, the tool particularly helped users who were less experienced with STPA find more UCAs. The tool also gave participants a higher confidence than their manual counterparts that they found all relevant UCAs, though several noted that doing so essentially relied on having suitably complete system models to use as input. Our results therefore suggest that the study participants do perceive a benefit from our approach.


\emph{\textbf{RQ2:}}
Our interview data suggest that the way FASR displays UCAs has downsides and can be improved. Like all formal methods tools, Fortis can produce a larger amount of output than is easily comprehended by users. FASR currently includes basic filtering, based on the number of UCAs per guideword; however, our data suggests that more sophisticated filtering (e.g., based on a notion of semantic diversity in UCAs) could have positive impact on usability in future iterations of the tool.

Additionally, participants noted that expanding our input language to include additional SysML diagram types (\eg, sequence or block diagrams) could provide more information to Fortis and enhance the specificity of the analysis. One member of the tool cohort explained that, while generating UCAs for STPA's third step is important, the main timesink / challenge is its fourth step: identifying loss scenarios, \ie, the ways the UCAs could occur given the system's implementation. The models we use as input do not contain the level of information necessary to automate this step, so extending the tool in this way would be non-trivial.

\subsubsection{Threats to Validity}
\label{sec:eval-challenges}
We mitigate the threats to validity in our study by focusing on exploratory research questions that are well-scoped, yet informative for the discussions above. For example, RQ1 is in terms of \textit{perceived} benefits; the data to answer this question is readily available in the interview transcripts, mitigating internal validity. On the other hand, RQ2 is in terms of the downsides and improvements that participants \textit{identify}, which is also readily available in the interview transcripts and mitigates external validity (\ie, generalization to other FASR users).


\section{Related Work}
\label{sec:related}

The use of formal methods for STPA has been explored in prior works.  Datkwa and Villani propose an approach that uses the UPPAAL model checker~\cite{behrmann2004tutorial} to check whether certain control actions could result in hazards over a formal model of a system~\cite{DAKWAT2018130}; the identification of UCAs in their approach is performed manually by a human analyst, while our tool generates such UCAs automatically. Integration of STPA with Event-B is another line of work in this direction~\cite{stpa-event-b-2023,colley2013stpa_eventb}, although their goal is to develop a safe-by-construction system design through refinement rather than automating parts of STPA. Abdulkhaleq \etal{} proposes an approach in which a software implementation is formally verified against safety requirements derived using STPA~\cite{ABDULKHALEQ20152}. In addition, alternative formal semantics for STPA have been proposed~\cite{thomas2013formalModelBasedSafetyReqs,stankovic-fstpa}, although their utility is demonstrated mainly by manual application to case studies.


Jung \etal{} describe a tool-supported process for extracting the \emph{process model} used by STPA's second step from a specification of a system's output variables and control actions. Those can then be converted to UCAs, though Fault Tree Analysis (FTA) is used to filter the control actions as the approach can produce too many UCAs for analysts without that additional step \cite{jung_formal_2022}. Petzold et al. describe \texttt{HotPASTA}, which is a tool that presents a number of improvements to STPA, including some amount of permutation-based UCA generation \cite{petzold_hot_2026}.


STPA has been embedded in various MBSA tools and processes. These include SysML~\cite{souza-stpa-sysml,ahlbrecht_model-based_2022}, AADL~\cite{camet-documentation,Procter2014} and the Capella tool set \cite{hetherington_stpa_2022}. In many cases, the actual work of performing STPA is done manually because, \eg, the tool's primary goal is traceability \cite{ahlbrecht_model-based_2022}, or report generation \cite{camet-documentation,Procter2014}, or supporting the use of system models when performing a manual STPA \cite{hetherington_stpa_2022}. de Souza \etal's work incorporates formal methods, though only for simulation after the analysis is complete \cite{souza-stpa-sysml}.

\section{Future Work}

FASR, like other automated MBSA tools, partially transforms an analytical problem (identifying hazards) into a modeling problem (creating sufficiently precise models). While manual STPA also relies on system models, they are typically much more abstract than the models used as input for FASR, and building these high-fidelity models of system behavior can be challenging and time consuming. We did not include model creation / editing tasks in our study due to the time required and to avoid further narrowing our participant pool, but we would like to analyze the time required for specifying these models in a future study.

Participants in our user study also identified several issues with the FASR tool itself; these also point towards potential next steps. We would like to explore clustering the outputs from Fortis in a way that is meaningful to users, and provides a clearer understanding of the underlying safety issue. We also plan on exploring alternative input formats; though whether to include additional SysML diagrams or switch to an alternate language like SysMLv2 is still under discussion. Finally, integrating our tool with a larger tool suite and / or tool-supported process, in particular one that addresses traceability, is something we are exploring.
\vspace{-1.5em}
\section*{Acknowledgments}
\vspace{-1em}
{\scriptsize
Copyright 2026 Carnegie Mellon University.

This material is based upon work supported by the Department of War under Air Force Contract No. FA8702-15-D-0002 with Carnegie Mellon University for the operation of the Software Engineering Institute, a federally funded research and development center.  

The opinions, findings, conclusions, and/or recommendations contained in this material are those of the author(s) and should not be construed as an official US Government position, policy, or decision, unless designated by other documentation.

[DISTRIBUTION STATEMENT A] This material has been approved for public release and unlimited distribution.  Please see Copyright notice for non-US Government use and distribution.

This work is licensed under a Creative Commons Attribution-NonCommercial 4.0 International License (https://creativecommons.org/licenses/by-nc/4.0/).  Requests for permission for non-licensed uses should be directed to the Software Engineering Institute at permission@sei.cmu.edu.

DM26-0561
}
\vspace{-1.4em}

\bibliographystyle{splncs04}
\bibliography{main.bib}

@techreport{leveson-thomas:BOOK18,
	title = {{STPA} {Handbook}},
	author = {Leveson, Nancy and Thomas, John},
	year = {2018},
	pages = {1--188},
}

@inproceedings{behrmann2004tutorial,
  author = {Behrmann, Gerd and David, Alexandre and Larsen, Kim G.},
  title = {A Tutorial on {Uppaal}},
  booktitle = {Formal Methods for the Design of Real-Time Systems},
  series = {LNCS},
  volume = {3185},
  pages = {200--236},
  publisher = {Springer, Berlin, Heidelberg},
  year = {2004}
}

@article{damerau-distance,
author = {Damerau, Fred J.},
title = {A technique for computer detection and correction of spelling errors},
year = {1964},
issue_date = {March 1964},
publisher = {Association for Computing Machinery},
address = {New York, NY, USA},
volume = {7},
number = {3},
issn = {0001-0782},
journal = {Commun. ACM},
month = mar,
pages = {171–176},
numpages = {6}
}

@INPROCEEDINGS{ltl,
  author={Pnueli, Amir},
  booktitle={18th Annual Symposium on Foundations of Computer Science (sfcs 1977)}, 
  title={The temporal logic of programs}, 
  year={1977},
  volume={},
  number={},
  pages={46-57},
}

@book{lamport2002specifying,
    author = {Lamport, Leslie},
    title = {Specifying Systems: The TLA+ Language and Tools for Hardware and Software Engineers},
    year = {2002},
    month = {6},
    publisher = {Addison-Wesley},
}

@inproceedings{fortis-fmcad23,
  author       = {Changjian Zhang and
                  Ian Dardik and
                  R{\^{o}}mulo Meira{-}G{\'{o}}es and
                  David Garlan and
                  Eunsuk Kang},
  title        = {Fortis: {A} Tool for Analysis and Repair of Robust Software Systems},
  booktitle    = {Formal Methods in Computer-Aided Design ({FMCAD})},
  pages        = {1--9},
  publisher    = {{IEEE}},
  year         = {2023}
}

@inproceedings{robustness-fse20,
author = {Zhang, Changjian and Garlan, David and Kang, Eunsuk},
title = {A Behavioral Notion of Robustness for Software Systems},
year = {2020},
isbn = {9781450370431},
booktitle = {Proceedings of ESEC/FSE},
pages = {1–12},
numpages = {12},
}

@inproceedings{colley2013stpa_eventb,
  author    = {John Colley and Michael Butler},
  title     = {A Formal, Systematic Approach to STPA Using Event-B Refinement and Proof},
  booktitle = {Proceedings of the 21st Safety-Critical Systems Symposium},
  year      = {2013},
  month     = {February},
  address   = {Bristol, UK},
  organization = {Safety-Critical Systems Club},
}

@INPROCEEDINGS{stankovic-fstpa,
  author={Asare, Philip and Lach, John and Stankovic, John A.},
  booktitle={2013 ACM/IEEE International Conference on Cyber-Physical Systems (ICCPS)}, 
  title={FSTPA-I: A formal approach to hazard identification via system theoretic process analysis}, 
  year={2013},
  volume={},
  number={},
  pages={150-159},
}

@INPROCEEDINGS{souza-stpa-sysml,
  author={de Souza, Fellipe Guilherme Rey and de Melo Bezerra, Juliana and Hirata, Celso Massaki and de Saqui-Sannes, Pierre and Apvrille, Ludovic},
  booktitle={2020 IEEE International Systems Conference (SysCon)}, 
  title={Combining STPA with SysML Modeling}, 
  year={2020},
  volume={},
  number={},
  pages={1-8},
  }

@inproceedings{thomas2013formalModelBasedSafetyReqs,
  author    = {Thomas, John and Leveson, Nancy},
  title     = {Generating Formal Model-Based Safety Requirements for Complex, Software- and Human-Intensive Systems},
  booktitle = {Proceedings of the Twenty-first Safety-Critical Systems Symposium},
  year      = {2013},
  address   = {Bristol, United Kingdom},
  organization = {Safety-Critical Systems Club},
}

@article{ABDULKHALEQ20152,
title = {A Comprehensive Safety Engineering Approach for Software-Intensive Systems Based on STPA},
journal = {Procedia Engineering},
volume = {128},
pages = {2-11},
year = {2015},
note = {Proceedings of the 3rd European STAMP Workshop},
issn = {1877-7058},
author = {Asim Abdulkhaleq and Stefan Wagner and Nancy Leveson}
}

@InProceedings{stpa-event-b-2023,
author="Salehi Fathabadi, Asieh
and Snook, Colin
and Dghaym, Dana
and Hoang, Thai Son
and Alotaibi, Fahad
and Butler, Michael",
editor="Gl{\"a}sser, Uwe
and Creissac Campos, Jose
and M{\'e}ry, Dominique
and Palanque, Philippe",
title="Designing Critical Systems Using Hierarchical STPA and Event-B",
booktitle="Rigorous State-Based Methods",
year="2023",
publisher="Springer Nature Switzerland",
address="Cham",
pages="220--237",
isbn="978-3-031-33163-3"
}

@article{DAKWAT2018130,
title = {System safety assessment based on STPA and model checking},
journal = {Safety Science},
volume = {109},
pages = {130-143},
year = {2018},
issn = {0925-7535},
author = {Alheri Longji Dakwat and Emilia Villani}
}

@article{de_saqui-sannes_taxonomy_2022,
	title = {A {Taxonomy} of {MBSE} {Approaches} by {Languages}, {Tools} and {Methods}},
	volume = {10},
	issn = {2169-3536},
	urldate = {2026-02-03},
	journal = {IEEE Access},
	author = {De Saqui-Sannes, Pierre and Vingerhoeds, Rob A. and Garion, Christophe and Thirioux, Xavier},
	year = {2022},
	pages = {120936--120950},
}

@inproceedings{ahlbrecht_model-based_2022,
	title = {Model-{Based} {STPA}: {Towards} {Agile} {Safety}-{Guided} {Design} with {Formalization}},
	issn = {2687-8828},
	shorttitle = {Model-{Based} {STPA}},
	urldate = {2026-02-03},
	booktitle = {2022 {IEEE} {International} {Symposium} on {Systems} {Engineering} ({ISSE})},
	author = {Ahlbrecht, Alexander and Zaeske, Wanja and Durak, Umut},
	month = oct,
	year = {2022},
	note = {ISSN: 2687-8828},
	pages = {1--8},
}

@inproceedings{Procter2014,
	title = {An architecturally-integrated, systems-based hazard analysis for medical applications},
	isbn = {978-1-4799-5338-7},
	booktitle = {12th {ACM}/{IEEE} {International} {Conference} on {Methods} and {Models} for {System} {Design}, {MEMOCODE} 2014},
	author = {Procter, Sam and Hatcliff, John},
	year = {2014},
}

@techreport{omg-sysml:TECH24,
    author = {{OMG\textsuperscript{\textregistered} Systems Modeling}},
    title = {{OMG Systems Modeling Language}\texttrademark ({SysML}\textsuperscript{\textregistered}) Version 1.7},
    institution = {{OMG Standards Development Organization}},
	month = {January},
    year = 2024,
	number = {formal/24-01-07},
}

@techreport{omg-raaml:TECH24,
    title = {{Risk Analysis and Assessment Modeling Language (RAAML) Libraries and Profiles Version 1.1 Beta 2}},
    institution = {{OMG Standards Development Organization}},
	month = {March},
    year = 2024,
	number = {ptc/24-03-02},
}

@misc{cameo2024x,
    author = {{Dassault Syst\`{e}mes}},
    title = {{Cameo Enterprise Architecture}},
    url = {https://www.3ds.com/products/catia/no-magic/cameo-enterprise-architecture},
    note = {2024x},
    year = {2026},
    month = {January},
    date = {2026-01-31},
}

@misc{camet-documentation,
    author= {{Galois, Inc}},
    year  = {2025},
    title = {CAMET Tools},
    note  = {\url{https://tools.galois.com/camet/overview/camet-tools}, 
             Last accessed on 2026-03-03},
}

@article{jung_formal_2022,
	title = {A formal approach to support the identification of unsafe control actions of {STPA} for nuclear protection systems},
	volume = {54},
	issn = {1738-5733},
	number = {5},
	urldate = {2025-04-22},
	journal = {Nuclear Engineering and Technology},
	author = {Jung, Sejin and Heo, Yoona and Yoo, Junbeom},
	month = may,
	year = {2022},
	pages = {1635--1643},
}

@inproceedings{hetherington_stpa_2022,
	address = {Toulouse, France},
	title = {{STPA} {Analysis} of {Automotive} {Safety} {Using} {Arcadia} and {Capella}},
	language = {en},
	booktitle = {{ERTS} 2022},
	publisher = {HAL Open Science},
	author = {Hetherington, David and Roques, Pascal},
	month = jun,
	year = {2022},
	pages = {191--200}
}

@book{ericson:BOOK16,
	address = {Fredericksburg, Virginia, United States of America},
	edition = {Second},
	title = {Hazard {Analysis} {Techniques} for {System} {Safety}},
	isbn = {978-1-118-94038-9},
	publisher = {John Wiley \& Sons, Inc.},
	author = {Ericson II, Clifton A.},
	year = {2016},
}

@inproceedings{petzold_hot_2026,
	address = {Cham},
	title = {Hot {PASTA}: {Improved} {Pragmatics} for {System}-{Theoretic} {Process} {Analysis}},
	isbn = {978-3-032-01241-8},
	shorttitle = {Hot {PASTA}},
	language = {en},
	booktitle = {Computer {Safety}, {Reliability}, and {Security}},
	publisher = {Springer Nature Switzerland},
	author = {Petzold, Jette and von Hanxleden, Reinhard},
	editor = {Gallina, Barbara and Törngren, Martin and Bitsch, Friedemann},
	year = {2026},
	pages = {160--174},
}

\end{document}